\documentclass[reprint,amsmath,amssymb,aps,prl,superscriptaddress,longbibliography]{revtex4-1}

\usepackage{bm}
\newcommand{\ii}{\mathrm{i}}
\newcommand{\ee}{\mathrm{e}}
\newcommand{\vk}{{\bf{k}}}
\newcommand{\vkp}{{\bf{k}^\prime}}
\newcommand{\eb}{\widetilde{\epsilon}_{\rm b}}
\newcommand{\ebg}{{\epsilon_{\rm bg}}}

\usepackage{graphicx}

\usepackage[
	colorlinks,
	breaklinks,]{hyperref}

\begin{document}

\title{
Topological two-band electron-hole superconductors with \textit{d}-wave symmetry: \\
Absence of Dirac quasiparticle annihilation in 
magic-angle twisted trilayer graphene
}

\author{Senne Van Loon}
\email{Senne.Van\_Loon@colostate.edu}
\affiliation{School of Physics, Georgia Institute of Technology, 
    Atlanta, Georgia 30332, USA}
\altaffiliation[Current address (SVL): ]{
    Department of Atmospheric Science, 
    Colorado State University, Fort Collins, Colorado 80521, USA}
\author{C. A. R. S\'a de Melo}
\affiliation{School of Physics, Georgia Institute of Technology, 
    Atlanta, Georgia 30332, USA}

\date{\today}

\begin{abstract}
We discuss a two-band model for two-dimensional superconductors with electron
and hole bands separated by an energy gap and singlet $d$-wave pairing in each
band. This type of model exhibits a V-shaped to U-shaped transition in the
density of the states of the superconductor, and was phenomenologically used as
a possible interpretation of recent tunneling experiments in magic-angle twisted
trilayer graphene (MATTG)~[Kim {\it et al.}, Nature {\bf 606}, 494-500 (2022)]. Performing a microscopic
investigation, we find that such a qualitative difference in behavior occurs
when the electron and hole chemical potentials change, leading to topological
quantum phase transitions (TQPTs) between gapless and gapped $d$-wave
superconducting states, due to the annihilation of chiral Dirac fermions at the
phase boundaries. This transition requires the vanishing of the coherence peaks
in the density of states at zero energy when the phase boundary is crossed, but
this is not seen in the experimental data of Kim {\it et al.} (2022). We also show that
direct thermodynamic signatures of these topological quantum phase transitions
arise in the theoretical compressibility, which exhibits logarithmic
singularities at the transition points. Measurements of the
compressibility may illuminate the interpretation of the
experimental data of Kim {\it et al.} (2022) and provide additional information
about topological quantum phase transitions in the superconducting state of
MATTG. Based on our analysis, we are led to conclude that the V-shaped to
U-shaped transition observed is not related to annihilation of
Dirac fermion quasiparticles and its associated TQPTs, but is possibly connected
to a change in symmetry of the order parameter from a nodal to a non-nodal superconducting phase.
\end{abstract}

\maketitle
%

In the search for high temperature superconductors, more and more evidence
points to two-dimensional (2D) structures as good candidates, but their
theoretical understanding remains limited, because many of these materials have
a complex structure. Specifically, multilayer graphene has been put forward as a
promising testbed for unconventional superconductivity~\cite{Cao2018, Oh2021,
Park2021, Kim2022, Zhang2022, Park2022,Alicea2023,Xie2024}. When the graphene
layers are twisted by a certain `magic' angle, the emerging moir\'e superlattice
allows for a superconducting phase that can be tuned with electric fields.
Experiments in magic-angle twisted bilayer~\cite{Cao2018,Oh2021} and
trilayer~\cite{Kim2021, Park2021,Kim2022} graphene (MATBG and MATTG) point to
the existence of the evolution of Bardeen-Cooper-Schrieffer (BCS) to Bose
superconductivity with a nodal order parameter~\cite{Senthil2022} as a function
of carrier density.

The BCS-Bose evolution has been studied theoretically~\cite{Nozieres1985,
SadeMelo1993} and experimentally~\cite{Zwierlein2005, Lompe2021Science} in the
context of $s$-wave ultracold atoms, where single band systems are easily
created and interactions are continuously changed.  Compared to the smooth
crossover in $s$-wave systems, interacting fermions with higher angular momentum
pairing undergo topological quantum phase transitions (TQPT) when interaction or
density are tuned~\cite{Duncan2000,
Botelho2005dwave,Botelho2005pwave,SaDeMelo2024}. Recent work has suggested that
high-$T_c$ cuprates may exhibit such evolution~\cite{Harrison2022}, while
another indicates that experimental evidence is
absent~\cite{Kivelson-Yu-He-2022}. Albeit the experimental success with $s$-wave
systems, observing tunable $p$- or $d$-wave superfluidity in cold-atom
experiments has been a challenge~\cite{Jin2003,Gorceix2009, Thywissen2023}, so
the advent of twisted graphene structures created an exciting new avenue to
study such TQPT in superconductors~\cite{Alicea2023}. 

In particular, recent experiments in MATTG~\cite{Kim2022} found evidence of a
phase transition between gapless and  gapped superconducting phases, as
indicated by a dramatic change from a V-shaped to a U-shaped profile in the
tunneling conductance, which measures the density of states (DOS). As suggested
in the experimental work of the Caltech group~\cite{Kim2022}, a possible
explanation for this behavior is the presence of a nodal $d$-wave order
parameter in a system with an electron and a hole band, where Andreev reflection
spectroscopy could be used to detect the proposed TQPT~\cite{Alicea2023}. These
recent works~\cite{Kim2022,Alicea2023} have relied on earlier theoretical
predictions of TQPT in one-band superconductors and
superfluids~\cite{Duncan2000, Botelho2005dwave, Botelho2005pwave}.

In this paper, we provide a deeper microscopic understanding of two-band
$d$-wave superconductors with electron and hole bands and tunable chemical
potentials or densities. Our model presents a rich phase diagram with distinct
superconducting phases that are topological in nature. The $d$-wave symmetry of
the pairing interaction allows for multiple TQPT, distinguished by the presence
of a gap in the quasiparticle spectrum, which emerges via the annihilation of
chiral Dirac fermions. We show that these transitions are robust to the breaking
of particle-hole symmetry and describe a transition from V-shaped to U-shaped
density of states. However, our microscopic conclusions are in contraposition to
the phenomenological interpretation given in experiments~\cite{Kim2022}, using
the same model, because such a TQPT requires the disappearance of coherence
peaks at the phase boundary, due to annihilation of chiral Dirac fermions, a
behavior that is not seen in experiments~\cite{Kim2022}. Furthermore, we show
that a direct thermodynamic signature of the TQPT must be seen in the
compressibility, which exhibits a logarithmic singularity at the transition
point due to the annihilation of chiral Dirac fermions. Recent compressibility
measurements~\cite{Xie2024} in the normal state of MATTG can potentially be
performed in the superconducting state. Such experiments may provide
thermodynamic information about the V-shaped to U-shaped transition and its
relation to the annihilation of Dirac fermion quasiparticles during the BCS-Bose
evolution.

\begin{figure}
    \centering 
    \includegraphics[width = 8.6cm]{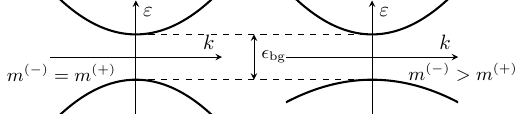}
    \caption{
        Energy bands of the free fermions with (right) and without (left) mass
        imbalance.
    }
    \label{fig:KinEn}
\end{figure}

\textit{Effective action: }
In general, the action for a two-band system of spin-1/2 fermions is 
\begin{align}
    S &= \beta \sum_{k,\sigma=\uparrow,\downarrow} \sum_{j=\pm}
        \bar{\psi}^{(j)}_{k,\sigma} 
            (-\ii\hbar\omega_n + \xi_{\vk}^{(j)})
        \psi_{k,\sigma}^{(j)} \\
    &\kern-1em - \frac{\beta}{L^2}\!\!\!\sum_{k,k^\prime,q} \sum_{i,j=\pm} \!\!
        V_{ij}(\vk,\vkp)
        \bar{\psi}^{(i)}_{k+\frac{q}{2},\uparrow} 
        \bar{\psi}^{(i)}_{-k+\frac{q}{2},\downarrow} 
        \psi^{(j)}_{-k^\prime+\frac{q}{2},\downarrow} 
        \psi^{(j)}_{k^\prime+\frac{q}{2},\uparrow}, \notag
\end{align}
where the fields $\psi_{k,\sigma}^{(j)}$ represent fermions of spin $\sigma$ in
band $j$. We use the notation $k = (\vk,\omega_n)$, with $\vk$ being momentum,
$\omega_n = \pi (2n+1)/\beta$ fermionic Matsubara frequencies, and $\beta =
\hbar/k_{\rm B} T$. From this action, the partition sum is defined as
$\mathcal{Z} = \int \mathcal{D}\bar{\psi}^{(+)} \mathcal{D}\psi^{(+)}
\mathcal{D}\bar{\psi}^{(-)} \mathcal{D}\psi^{(-)} e^{-S/\hbar}$. We study a
system with an electron (particle) and a hole band separated by a gap $\ebg$
(see Fig.~\ref{fig:KinEn}), and denote them by $j = +$ and $j = -$ respectively.
The kinetic energy of the bands is given by $\epsilon_{\vk}^{(\pm)} = \pm
(\hbar^2\vk^2/2m^{(\pm)} +\ebg/2)$, and $\xi_\vk^{(\pm)} =
\epsilon_{\vk}^{(\pm)} \mp \mu^{(\pm)}$, where the chemical potential
$\mu^{(+)}$ $(\mu^{(-)})$ for the electron (hole) is measured with respect to
the bottom (top) of the electron (hole) band $+\ebg/2$ $(-\ebg/2)$. We allow for
both intraband and interband coupling, and write the pairing interaction
$V_{ij}(\vk,\vkp)= \lambda_{ij} \Gamma_\vk \Gamma_\vkp$ in separable form. The
symmetry factors $\Gamma_\vk$ characterize the angular momentum $\ell$ of the
pairing channel~\cite{Duncan2000,Botelho2005pwave}, which we take to be of the
$d$-wave type $(\ell = 2)$ in two dimensions, such that $\Gamma_\vk =
\gamma_{\vk} \cos 2\phi$ and $\gamma_{\vk} = (\vert\vk\vert/k_1)^2 (1+
\vert\vk\vert/k_0)^{-5/2}$. Here, $k_0$ and $k_1$ set the scales at high and low
$\vert {\bf k}\vert$ respectively, with $R_0 \sim k_0^{-1}$ being the effective
interaction range. We present our results in units of $\epsilon_0 =
\hbar^2k_0^2/2m^{(+)}$ and set $k_1 = k_0$ without loss of generality, as $k_1$
only scales the interaction prefactors $\lambda_{ij}$.

Superconductivity emerges through the complex Hubbard-Stratonovich pair fields
$\Phi_r^{(\pm)} = |\Delta^{(\pm)}| (1 + \phi_r^{(\pm)}) \ee^{\ii
\theta_r^{(\pm)}}$, where $\Delta^{(\pm)}$ is order parameter,  while
$\phi_r^{(\pm)}$ is the modulus and $\theta_r^{(\pm)}$ is the phase fluctuation
in each band $(\pm)$. For slowly varying bosonic fields with respect to
fermionic degrees of freedom, the action becomes
\begin{align}
    S_{\rm HS} =& \int \frac{d^3r}{L^2} \bigg[
        \sum_{k} \bar{\eta}_{k} G^{-1}_{k}(r) \eta_{k}
        +\sum_{i,j=\pm} (\Phi^{(i)}_r)^\dagger g_{ij} \Phi^{(j)}_r \notag \\
        &\kern3em +\sum_{\vk,j=\pm} j\Big(
            \xi_{\vk}^{(j)} + \chi_{\vk}^{(j)}(r) + \gamma_{\vk}^{(j)}(r)
        \Big)
    \bigg].
\end{align}
We use the notation $r=({\bf{r}},\tau)$, $\int d^3r = \int_0^{\beta} d\tau
\int_{L^2} d^2{\bf{r}}$, define the effective coupling $g_{ij} =
ijL^2\lambda_{-j,-i}/(\lambda_{++}\lambda_{--}-\lambda_{+-}\lambda_{-+})$, and
the functions $\gamma_\vk^{(\pm)}(r) = \ii\hbar\partial_\tau \theta_r^{(\pm)}/2
+ (\hbar^2/8m) ({\boldsymbol\nabla}\theta_r^{(\pm)} )^2$ and
$\chi_\vk^{(\pm)}(r) = (\hbar^2/4m) \ii\nabla^2{\theta_r^{(\pm)}}-(\hbar^2/2m)
{\boldsymbol\nabla}{\theta_r^{(\pm)}} \cdot \vk$. The fermionic fields are
represented by the Nambu spinors $\bar{\eta}_k = (\bar{\psi}^{(+)}_{k,\uparrow},
\psi^{(+)}_{-k,\downarrow}, \bar{\psi}^{(-)}_{k,\uparrow},
\psi^{(-)}_{-k,\downarrow})$, and the inverse quasiparticle propagator is a
blockdiagonal $4\times 4$ matrix $G^{-1} = {\rm
diag}[(G^{(+)})^{-1},(G^{(-)})^{-1}]$ that can be separated into saddle point
and fluctuation parts $(G^{(\pm)})^{-1} = (G_0^{(\pm)})^{-1} + F^{(\pm)}$, where
\begin{align}
    (G_0^{(\pm)})^{-1} =& 
    \begin{pmatrix}
        -\ii \hbar \omega_n + \xi_{\vk}^{(\pm)} 
        & - |\Delta_{\vk}^{(\pm)}| \\
        -|\Delta_{\vk}^{(\pm)}| 
        & -\ii \hbar \omega_n - \xi_{\vk}^{(\pm)}
    \end{pmatrix}, \label{quasiProp}\\
    F^{(\pm)} =& 
    \begin{pmatrix}
        \chi_{\vk}^{(j)}(r) + \gamma_{\vk}^{(j)}(r) 
        & - |\Delta_{\vk}^{(\pm)}| \phi_r^{(\pm)} \\
        -|\Delta_{\vk}^{(\pm)}| \phi_r^{(\pm)} 
        & \chi_{\vk}^{(j)}(r)- \gamma_{\vk}^{(j)}(r)
    \end{pmatrix},
\end{align}
with $\Delta_{\vk}^{(\pm)} \equiv \Delta^{(\pm)}\Gamma_{\vk}$. 

Integrating the fermions leads to 
the effective action
\begin{align}
S_{\rm eff} & = \sum_{j=\pm} \bigg[ 
 S_{\rm sp}^{(j)} + \frac{1}{2} \int d^3r \bigg\{
 A^{(j)} (\hbar\partial_\tau\theta_r^{(j)})^2 
  \notag \\
&
+\rho_{s}^{(j)} (\nabla{\theta_r^{(j)}})^2 
+
C^{(j)} (\phi_r^{(j)})^2 
 + \ii D^{(j)} \phi_r^{(j)} \hbar\partial_\tau\theta_r^{(j)}
\bigg\}
\bigg] \notag \\
&+ 2 g_{+-} \int \frac{d^3r}{L^2} |\Phi_r^{(+)}||\Phi_r^{(-)}|
 \cos(\theta_r^{(+)}-\theta_r^{(-)}),
\label{eq:effAction}
\end{align}
The first term in~Eq.~\eqref{eq:effAction} describes the saddle point result
\begin{align}
    &\frac{1}{\beta}S_{\rm sp}^{(\pm)} = 
        g_{\pm\pm} |\Delta^{(\pm)}|^2  \\
        &\quad+\sum_{\vk} \bigg[ 
            \pm \xi_{\vk}^{(\pm)} - E_{\vk}^{(\pm)} - 2 k_{\rm B} T\log\big(
                1+ \ee^{-E_{\vk}^{(\pm)}/k_{\rm B} T}
            \big)
        \bigg], \notag
\end{align}
where $E_\vk^{(\pm)} = \sqrt{(\xi_{\vk}^{(\pm)})^2 + |\Delta_{\vk}^{(\pm)}|^2}$
are the quasiparticle energies. The second term is the effective action for each
band describing phase $\theta_{r}^{(\pm)}$ and modulus fluctuations
$\phi_r^{(\pm)}$, where $A^{(\pm)}$ is related to the compressibility at fixed
temperature and order parameter; $\rho^{(\pm)}$ represents the superfluid
density; $C^{(\pm)}$ describes the modulus fluctuations; and $D^{(\pm)}$
represents the coupling between modulus and phase fluctuations. The last line
in~Eq.~\eqref{eq:effAction} describes the Josephson coupling between the bands.

We consider the saddle point approximation at $T=0$, where the thermodynamic
potential $\Omega = -k_{\rm B} T\ln \mathcal{Z}$ reduces to
\begin{align}
    \Omega_{\rm sp} =& \sum_{j=\pm} \Big[
        g_{jj} |\Delta^{(j)}|^2 
        + \sum_{\vk} \Big(  j \xi_{\vk}^{(j)} - E_{\vk}^{(j)} \Big)
    \Big] \notag \\
    &+ 2 g_{+-}  |\Delta^{(+)}|  |\Delta^{(-)}| \cos \theta^\prime,
\end{align}
with $\theta^\prime$ being the relative phase difference between the order
parameters in each band. The order parameters are found from the condition
$\partial\Omega_{\rm sp}/\partial|\Delta^{(\pm)}| = 0$, leading to two coupled
equations
\begin{equation}
    g_{\pm\pm}|\Delta^{(\pm)}| + g_{+-} |\Delta^{(\mp)}| \cos \theta^\prime = \sum_\vk 
        \frac{|\Delta^{(\pm)}| \Gamma_{\vk}^2}{2E_{\vk}^{(\pm)}}.
    \label{eq:OPeq}
\end{equation}
The order parameters $\Delta^{(\pm)}$ are functions of $\mu^{(+)}$ and
$\mu^{(-)}$ or equivalently of the total chemical potential $\mu =
(\mu^{(+)}+\mu^{(-)})/2$ and the imbalance $\zeta = (\mu^{(+)}-\mu^{(-)})/2$. To
relate $\vert\Delta^{(\pm)}\vert$ to electron $n^{(+)}$ and hole $n^{(-)}$
densities, we need to obtain the equations of state (number equations) for the
total density $n = n^{(+)} + n^{(-)}$ obtained from  $n = - \frac{1}{L^2}
\frac{\partial\Omega}{\partial \mu}$ and for the density difference $\delta n =
n^{(+)} - n^{(-)}$ written as $\delta n = - \frac{1}{L^2}
\frac{\partial\Omega}{\partial \zeta}$, where $\Omega$ is the thermodynamic
potential. Using the saddle point thermodynamic potential $\Omega _{\rm sp}$ to
obtain the equation of state (number equation) is not sufficient, because
quantum fluctuations are quantitatively important in 2D, specially in the Bose
regime~\cite{VanLoon2023}. Therefore, we show results in terms of chemical
potentials rather than carrier densities, and leave a complete analysis of the
equation of state including fluctuations for a later opportunity.

\begin{figure}
\centering 
\includegraphics{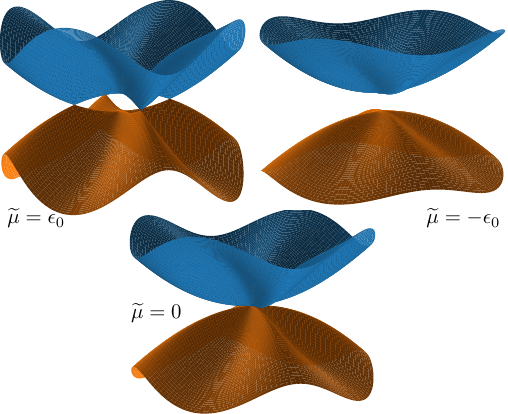}
\caption{
Quasiparticle spectrum for different values of the total chemical potential with
$\eb = -4\epsilon_0$. Here, $\vert\Delta^{(+)}\vert=\vert\Delta^{(-)}\vert$,
$\zeta = 0$ and $m^{(+)} = m^{(-)}$, such that $E_{\vk}^{(+)}=E_{\vk}^{(-)}$. 
}
\label{fig:QuasiEn_balanced}
\end{figure}

\textit{Topological quantum phase transition: }
The chemical potentials ${\widetilde\mu}^{(\pm)} = \mu^{(\pm)} - \epsilon_{\rm
bg}/2$ play an important role in determining the phase diagrams of our system,
as can be seen in the quasiparticle spectrum. In the most general case, four
quasiparticle/quasihole (qp-qh) bands are present, two for the electrons $\pm
E_{\vk}^{(+)}$ and two for the holes $\pm E_{\vk}^{(-)}$. Due to the nature of
the $d$-wave interactions, TQPT of the Lifshitz type~\cite{Duncan2000,
Botelho2005dwave} occur at ${\widetilde\mu}^{(\pm)}  = 0$, when the
quasiparticle energies go to zero at $|\vk| = 0$. For ${\widetilde\mu}^{(\pm)} >
0$, the quasiparticle spectrum is gapless, with $E_{\vk}^{(\pm)} = 0$ at
$\hbar|\vk| = \sqrt{2m^{(\pm)}\widetilde\mu^{(\pm)}} \equiv \hbar
k_{\mu}^{(\pm)}$ and $\phi = \pi/4 + n \pi/2$. There are four chiral Dirac
quasiparticles in the electron band and four in the hole band with topological
indices (winding numbers) 
\begin{equation}
N_w^{(\pm)} = 
\frac{1}{2\pi} \oint_C d \ell ~{\hat {\bf z}}
\cdot {\bf m}_{\pm} \times 
\frac{d {\bf m}_{\pm}}{d\ell},
\end{equation}
where ${\bf m}_{\pm} = (\xi_{\bf k}^{(\pm)}, \vert \Delta_{\bf k}^{(\pm)} \vert
)/E_{\bf k}^{(\pm)}$, and the path $C$ encircles each Dirac point separately.
Notice that $N_w^{(\pm)}$ alternates between $\mp1$ $(\phi = \pi/4, 5\pi/4)$ and
$\pm1$ $(\phi = 3\pi/4, 7\pi/4)$ as the Dirac points are circled
counterclockwise with increasing values of $\phi \in [0, 2\pi]$. The net
chirality of the system is zero, as there is no global chiral symmetry breaking.
However, the annihilation of these chiral Dirac quasiparticles at ${\widetilde
\mu}^{(\pm)} = 0$ defines TQPT in momentum space. For ${\widetilde \mu}^{(\pm)}<
0$, the spectrum is fully gapped, with $\min_{\vk} \{E_\vk^{(\pm)}\} = \vert
{\widetilde \mu}^{(\pm)} \vert$ at $|\vk| = 0$, and $N_w^{(\pm)}$ is always
zero, as there are no chiral Dirac quasiparticles. An example of the TQPT is
visible in Fig.~\ref{fig:QuasiEn_balanced}, where the quasiparticle energies are
shown in the simplest case where $E_{\vk}^{(+)} = E_{\vk}^{(-)}$.

Measurements of the density of states (DOS) via tunelling conductance in
MATTG~\cite{Kim2022} revealed a remarkable V-shaped to U-shaped transition for
sufficiently low-carrier densities as the chemical potential is tuned. Below, we
show that such transition naturally occurs in our model and is a direct
consequence of the chemical potentials ${\widetilde\mu}^{(\pm)} = \mu^{(\pm)} -
\epsilon_{\rm bg}/2$ being the relevant small energy scale in the problem. For
sufficiently weak interactions or high carrier densities and
$\widetilde{\mu}^{(\pm)} > 0$, $E_{\bf k}^{(\pm)}$ has a local maximum at ${\bf
k = 0}$ and saddle points at $\phi = n\pi/2$ and $|\vk|<k_\mu^{(\pm)}$ leading
to four peaks in the DOS at low energies, two on the positive and two on the
negative side, when ${\widetilde \mu}^{(\pm)} \sim \vert \Delta^{(\pm)} \vert
\gamma_{k_{\mu}^{(\pm)}}$. For strong interactions or low carrier densities and
$\widetilde{\mu}^{(\pm)} > 0$, the saddle points at nonzero ${\bf k}$ become
indistinguishable from the maximum at ${\bf k} = {\bf 0}$ producing two peaks in
the DOS at low energies, one on the positive and one on the negative side, when
${\widetilde \mu}^{(\pm)}\ll\vert \Delta^{(\pm)} \vert
\gamma_{k_{\mu}^{(\pm)}}$. A two-peak structure in the V-shaped phase is seen in
the MATTG DOS experiments~\cite{Kim2022}, suggesting that this system is in the
strong interaction regime.

The TQPT described above are controlled only by the chemical potentials
$\mu^{(\pm)}$, such that specific values of the masses $m^{(\pm)}$, interactions
$g_{++}, g_{--}, g_{+-}$, or order parameters $\vert \Delta^{(\pm)}\vert$ do not
matter. However, the transition from V-shaped to U-shaped in the low-energy DOS,
with two peaks only, requires that the interactions are sufficiently strong to
eliminate saddle points at nonzero ${\bf k}$ in $E_{\bf k}^{(\pm)}$, that is, it
requires that ${\widetilde \mu}^{(\pm)}\ll\vert \Delta^{(\pm)} \vert
\gamma_{k_{\mu}^{(\pm)}}$. To describe TQPT controlled by $\mu^{(\pm)}$, the
simplest choices of parameters are intraband interactions $g_{++} = g_{--}$
$(\lambda_{++} = \lambda_{--})$, band masses $m^{(+)} = m^{(-)}$, and small
Josephson coupling $g_{+-}$. Thus, in Eq.~\eqref{eq:OPeq}, we set $\vert
\Delta^{(+)} \vert = \vert \Delta^{(-)} \vert \equiv \vert \Delta \vert$ and
$g_{+-} \to 0$. Finally, we fix the interaction by using the relation between
the two-body binding energy ${\widetilde \epsilon}_{\rm b}^{(\pm)} =
\epsilon_{\rm b}^{(\pm)} - \epsilon_{\rm bg}$ with respect to $\epsilon_{\rm
bg}$ and the interaction parameters $L^2/\lambda_{\pm\pm} = \sum_\vk
\vert\Gamma_\vk\vert^2/(\hbar^2\vk^2/m^{(\pm)}-
{\widetilde\epsilon}_\mathrm{b}^{(\pm)})$. When the electron and hole masses and
the electron-electron and hole-hole interactions are the same, that is, $m^{(+)}
= m^{(-)}$ and $\lambda_{++} = \lambda_{--}$, the binding energies ${\widetilde
\epsilon}_{\rm b}^{(+)} = {\widetilde\epsilon}_{\rm b}^{(-)} =
{\widetilde\epsilon}_{\rm b}$ are also identical. 

\begin{figure}
\centering
\includegraphics[width = 8.6cm]{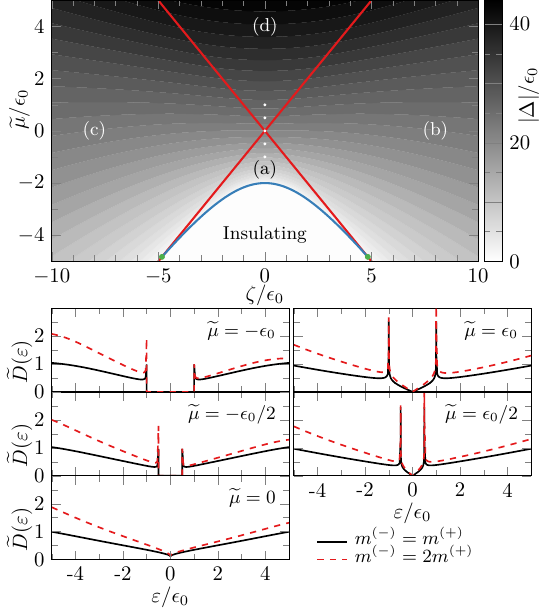}
\caption{
(Top) Density plot of $|\Delta|$ vs $\zeta/\epsilon_0$ and
$\widetilde{\mu}/\epsilon_0 = (\mu -\ebg/2)/\epsilon_0$ for $m^{(-)}=m^{(+)}$
and ${\widetilde \epsilon}_{\rm b} = -4\epsilon_0$. The red lines separate the
different superfluid phases, while the blue line shows the transition to an
insulating phase.  
Notice that $\vert \Delta \vert = 0$ only within the insulating region. The
green dots show critical values for $\vert \zeta_c\vert$ above which no fully
gapped superconductivity is possible. In (a) both the electron and hole sectors
are gapped (Bose-like), in (b) the electron sector is gapless (BCS-like), while
the hole sector is gapped (Bose-like), in (c) the electron sector is gapped
(Bose-like), while the hole sector is gapless (BCS-like), and in (d) both the
electron and hole sectors are gapless (BCS-like). The white dots show the values
of $\mu$ and $\zeta$ used in the bottom panel. (Bottom) The dimensionless
density of states $\widetilde{D}(\varepsilon) = \epsilon_0 D{(\varepsilon) }$
vs. $\varepsilon/\epsilon_0$ for $\zeta = 0$, and various $\widetilde{\mu}$ with
${\widetilde \epsilon}_{\rm b} = -4\epsilon_0$.
}
\label{fig:EoS}
\end{figure}
%

\textit{Phase diagram: }
On the top panel of Fig.~\ref{fig:EoS}, the phase diagram and $|\Delta|$ are
shown in the ${\widetilde\mu}$-$\zeta$ plane for ${\widetilde \epsilon}_{\rm b}
= -4 \epsilon_0$ ($\lambda_{\pm\pm}\simeq 121.9 \epsilon_0/k_0^2$), which
guarantees strong coupling in the sense of ${\widetilde \mu}^{(\pm)}\ll\vert
\Delta^{(\pm)} \vert \gamma_{k_{\mu}^{(\pm)}}$. As seen below, this choice of
$\eb$ gives a two-peaked DOS at low energies with a clear V-shaped to U-shaped
transition. There are five different phases depending on the chemical potentials
of the electrons and holes. Each quasiparticle/quasihole band undergoes a
Lifshitz transition when ${\widetilde \mu}^{(\pm)} = 0$, or equivalently when
${\widetilde \mu} = \pm \zeta$, shown as red lines in the phase diagram.  When
both ${\widetilde\mu}^{(\pm)} < 0$, the quasiparticle bands are fully gapped,
with both electron and hole sectors behaving as Bose-like superconductors [phase
(a)]. Conversely, when ${\widetilde\mu}^{(\pm)} > 0$, the quasiparticle bands
are gapless, with both electron and hole sectors behaving as BCS-like
superconductors [phase (d)]. When ${\widetilde \mu}^{(+)} > 0$ and ${\widetilde
\mu}^{(-)} < 0$, the quasiparticles for the electron sector are gapless
(BCS-like), while the quasiparticles for the hole sector are fully gapped
(Bose-like) [phase (b)]. When ${\widetilde\mu}^{(+)} < 0$ and ${\widetilde
\mu}^{(-)} > 0$, the quasiparticles for the electron sector are fully gapped
(Bose-like), while the quasiparticles for the hole sector are gapless (BCS-like)
[phase (c)]. The transition line into a normal (insulating) state with $|\Delta|
= 0$, where no charge carriers are available, is shown in blue on the top panel
of Fig.~\ref{fig:EoS}. Note that an insulating phase is only present when both
${\widetilde\mu}^{(\pm)} < 0$, or ${\widetilde \mu} < -|\zeta|$. Outside of this
region, the right hand side of Eq.~\eqref{eq:OPeq} with $|\Delta|=0$ contains an
imaginary part and a solution can never be found. Thus, there is a critical
value of the imbalance $|\zeta_c|$ (shown as green dots in Fig.~\ref{fig:EoS})
indicating the largest value of $|\zeta|$ where fully gapped superfluidity is
possible. For $|\zeta| > |\zeta_c|$ and ${\widetilde \mu} < -|\zeta|$, the
system is always in the insulating state. 

For $m^{(+)}=m^{(-)}$, the phase diagram is completely symmetric around
$\zeta=0$, reflecting the particle-hole symmetry in the energy spectrum.
However, for $m^{(+)}\neq m^{(-)}$ or $\lambda_{++}\neq \lambda_{--}$,
particle-hole symmetry is broken, but the red phase boundaries remain the same,
while the blue boundary becomes asymmetric with respect to $\zeta$. 

\textit{Density of states: }
The DOS for electrons or holes $D^{(\pm)}(\varepsilon) = -\sum_{\vk} {\rm Im}
[(G_{0}^{(\pm)}(\vk, \varepsilon + \ii 0^+))_{11}]/\pi$ is found by analytically
continuing $\ii \hbar\omega_n \rightarrow \varepsilon + \ii 0^+$ in the
quasiparticle propagator defined in Eq.~\eqref{quasiProp}. The total density of
states is the sum
\begin{align}
    D(\varepsilon) =& \!\!\sum_{\vk,j=\pm} \Big[
          \big(U_{\vk}^{(j)}\big)^2 \delta\big( \varepsilon - E_{\vk}^{(j)}\big) 
        + \big(V_{\vk}^{(j)}\big)^2 \delta\big(\varepsilon + E_{\vk}^{(j)} \big)
    \Big],
\end{align}
where $U_\vk^{(\pm)} = \sqrt{(1 + \xi_\vk^{(\pm)}/E_\vk^{(\pm)})/2}$ and
$V_\vk^{(\pm)} = \sqrt{(1 - \xi_\vk^{(\pm)}/E_\vk^{(\pm)})/2}$ are the
Bogoliubov coefficients. 

The black lines in the bottom panel of Fig.~\ref{fig:EoS} show the DOS for
$m^{(+)} = m^{(-)}$. If also the chemical potentials of electrons and holes are
equal, $\mu^{(+)} = \mu^{(-)}$, the DOS is particle-hole symmetric. Only when
${\widetilde \mu}<-|\zeta|$ is the system fully gapped, and a U-shape can be
discerned in the DOS, with $D(\varepsilon)$ being zero within the gap between
the quasiparticle and quasihole energies. In this case, the electron and hole
sectors exhibit Bose-like superconductivity. When one or both of the
quasiparticle bands undergo a Lifshitz phase transition to the gapless state,
either the electron or hole sector or both exhibit BCS-like superconductivity,
and the DOS acquires a V-shape, only going to zero at $\varepsilon = 0$. For
$\zeta = 0$, the peaks at the gap edges for the U-shaped regime occur at
$\varepsilon = \pm \vert {\widetilde \mu} \vert$, and the two peaks in the
V-shaped region also occur at $\varepsilon = \pm \vert {\widetilde \mu} \vert$,
since the system is in the strongly interacting regime. A similar transition was
observed in the tunneling conductance of MATTG~\cite{Kim2022}, however the
disappearance of the coherence peaks at ${\widetilde \mu} = 0$ (See
Fig.~\ref{fig:EoS}) due to the annihilation of chiral Dirac fermions was not
seen experimentally~\cite{Kim2022}.

For more complex band structures with additional Fermi pockets and extra superconducting gaps, more coherence peaks would be present. Still, the coherence peaks associated with Dirac fermion quasiparticles must move towards each other as $\widetilde \mu \to 0$ to allow for their disappearance, considering that the annihilation of Dirac fermions is the cause of the topological transition. In contrast, the coherence peaks not associated with Dirac fermions would not disappear as ${\widetilde \mu}$ is changed. 
This is a spectroscopic manifestation of the topological requirement that Dirac fermions annihilate at the phase boundary, and must occur for any model that invokes the disappearance of Dirac fermion quasiparticles as the cause of the topological transition.

Particle-hole symmetry can be broken in multiple ways. When $\zeta \neq 0$,
three different phases emerge with changing ${\widetilde \mu}$, and two
asymmetric peaks are present at $\varepsilon = \mp\widetilde{\mu}^{(\pm)}$. When
$m^{(+)} \ne m^{(-)}$, particle-hole symmetry is broken even at $\zeta=0$, see
the dashed red lines in the bottom panel of Fig.~\ref{fig:EoS}. The DOS becomes
larger on the hole side, because the mass of the holes is larger $m^{(-)} = 2
m^{(+)}$. However, the largest Van Hove singularity switches from positive to
negative energies at the TPQT when crossing the phase boundary from BCS-like
[phase (d)] to Bose-like [phase (a)], a feature that is controlled by the
Bogoliubov coefficients, which act as the residues of the poles at $\varepsilon
= \pm E_{\bf k}^{(\pm)}$.

\begin{figure}
\centering 
\includegraphics[width = 8.6cm]{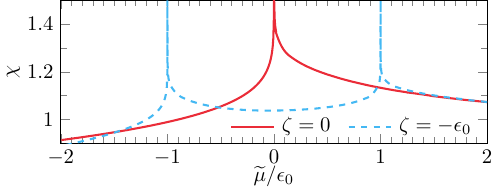}
\caption{
The dimensionless compressibility $\chi = n^2\kappa (2\pi \epsilon_0/k_0^2)$ as
a function of the total chemical potential $\widetilde{\mu}$ for different
values of the imbalance $\zeta$, within the superconducting phases, using $\eb =
-4\epsilon_0$ and $m^{(+)}=m^{(-)}.$ The location of topological quantum phase
transitions, due to the annihilation of Dirac fermion quasiparticles, are
indicated by the logarithmic singularities (peaks) in the solid red and dashed
blue lines.
}
\label{fig:compressibility}
\end{figure}

\textit{Compressibility: }
In Fig.~\ref{fig:compressibility}, we show the compressibility $\kappa = n^{-2}
d n/d \mu\vert_{T}$, with $n$ being the total density, as a thermodynamic
signature of TQPT. For $d$-wave pairing in 2D, logarithmic
singularities~\cite{Duncan2000, Botelho2005dwave} exist at the phase boundaries
where chiral Dirac fermions annihilate and gaps in the quasiparticle excitation
spectrum of the electron and/or hole sectors emerge. When changing the chemical
potential ${\widetilde \mu}$ following the line of constant $\zeta = 0$ (solid
red line in Fig.~\ref{fig:compressibility}), there is only one peak, because a
single TQPT occurs by going directly from region (a) to (d) (see dotted line in
Fig.~\ref{fig:EoS}). When changing ${\widetilde \mu}$ for fixed $\zeta \ne 0$
(dashed blue line in Fig.~\ref{fig:compressibility}) there are two peaks,
because two TQPTs occur: the first when going from (a) to (c), and the second
when going from (c) to (d) (see Fig.~\ref{fig:EoS}). Therefore, the number of
peaks (logarithmic singularities) represents the number of TQPT of the
Lifshitz-type that are crossed in the phase diagram of Fig.~\ref{fig:EoS}.

\textit{Conclusions:}
We presented a simple two-band model for two-dimensional superconductors with
electron and hole bands and singlet $d$-wave pairing in each band. We showed
that this model exhibits a TQPT that leads to a V-shaped to U-shaped transition
in the DOS of the superconductor similar to that found in tunneling experiments
on magic-angle twisted trilayer graphene~\cite{Kim2022}. However, topological
constraints require the annihilation of chiral Dirac fermions during  the
evolution from BCS-like to Bose-like superconductivity, leading to the
disappearance of coherence peaks at the transition point, an effect that is not
observed in experiments~\cite{Kim2022}. The lack of this observation is  a
sufficiently strong reason to discard the phenomenological interpretation that a
TQPT with $d$-wave symmetry occurs due to the annihilation of Dirac
fermions~\cite{Kim2022}. We also emphasized that a logarithmic singularity in
the compressibility must be observed whenever phase boundaries are crossed,
providing a thermodynamic test for the existence of such TQPT.  
Recently, the compressibility of MATTG was measured~\cite{Xie2024} in the normal
state; a similar measurement at low temperatures may provide additional
information about topological quantum phase transitions in the superconducting
phase. Based on our analysis, we are led to conclude that the V-shaped to
U-shaped transition observed~\cite{Kim2022} is not related to annihilation of
Dirac fermion quasiparticles and its associated TQPTs, but is possibly connected
to a change in symmetry of the order parameter from a nodal to a non-nodal superconducting phase.

\begin{acknowledgments}
{\it Acknowledgments:}
We thank the Belgian American Educational Foundation.
\end{acknowledgments}

\end{document}